\documentstyle[psfig]{mn}

\title[Search for Periodicities in X-ray Bursts of the Rapid Burster]
  {Search for Millisecond Periodicities in Type I X-ray Bursts of the
  Rapid Burster} 

\author[D. W. Fox et al.]{
            D.~W.~Fox,$^1$\footnote{derekfox@space.mit.edu}
            W.~H.~G.~Lewin,$^1$ R.~E.~Rutledge,$^2$ 
            E.~H.~Morgan,$^1$ R.~Guerriero,$^3$
 \newauthor L.~Bildsten,$^4$ M.~van~der~Klis,$^5$
            J.~van~Paradijs,$^{5,6}$\footnote{Deceased}
            C.~B.~Moore,$^7$ T.~Dotani,$^8$ K.~Asai$^8$\\
 $^1$Center for Space Research and Department of Physics,
       Massachusetts Institute of Technology, Cambridge, MA
       02139--4307, USA \\
 $^2$Space Radiation Laboratory, California Institute of
       Technology, MC 220-47, Pasadena, CA 91125, USA \\
 $^3$Department of Physics, United States Military
       Academy, West Point, NY 10996, USA \\
 $^4$Institute for Theoretical Physics and Department of
       Physics, University of California, Santa Barbara,
       Santa Barbara, CA 93106, USA \\
 $^5$Astronomical Institute `Anton Pannekoek' and Center
       for High-Energy Physics, Kruislaan 403, 1098 SJ Amsterdam, The
       Netherlands \\
 $^6$Department of Physics, University of Alabama
       Huntsville, Huntsville, AL 35899, USA \\ 
 $^7$Harvard-Smithsonian Center for Astrophysics, 60
       Garden Street, MS 78, Cambridge, MA 02138, USA \\
 $^8$Institute of Space and Astronautical Science,
        Sagamihara, Japan \\
 ~ \\
 $^\star${\rm derekfox@space.mit.edu} \\
 $^\dag${\rm Deceased}
}

\date{Accepted date. Received date.}
\pagerange{\pageref{firstpage}--\pageref{lastpage}}
\pubyear{2000}

\newcommand{\mxbrb}{\mbox{MXB~1730$-$335}}
\newcommand{\slowb}{\mbox{4U~1728$-$34}}

\newcommand{\ksxrb}{\mbox{KS~1731$-$260}}
\newcommand{\etal}{\mbox{et al.}}
\newcommand{\rxte}{\textit{RXTE}}
\newcommand{\rxtez}{\textit{RXTE\/}}

\newcommand{\ergcms}{ergs cm$^{-2}$ s$^{-1}$}
\newcommand{\ergsec}{ergs s$^{-1}$}
\newcommand{\ctsec}{c s$^{-1}$}

\newcommand{\twid}[1]{$\sim$#1}

\def\lesssim{\mathrel{\hbox{\rlap{\hbox{\lower4pt\hbox{$\sim$}}}\hbox{$<$}}}}
\def\gtrsim{\mathrel{\hbox{\rlap{\hbox{\lower4pt\hbox{$\sim$}}}\hbox{$>$}}}}

\newcommand{\Tfn}[1]{$^{\rm #1}$}
\newcommand{\multc}[1]{\multicolumn{1}{c}{#1}}
\newcommand{\multcc}[2]{\multicolumn{#1}{c}{#2}}
\newcommand{\markcite}[1]{#1}
\newcommand{\reference}[1]{\bibitem{x#1}}

\begin{document}
\label{firstpage}
\maketitle

\begin{abstract}
We have searched the rising portion of type~I X-ray bursts observed
from the Rapid Burster with the {\it Rossi X-ray Timing Explorer\/}
for the presence of periodicities. The 95 per cent confidence upper
limit on the average root-mean-square variation of near coherent
pulsations with a width of $\lesssim$1~Hz (in 60--2048~Hz) during the
first second of the bursts is $<$8.8 per cent. We find a possible
detection ($>$98 per cent significance) at 306.5 Hz.
\end{abstract}



\section{Introduction}
The past four years have seen dramatic advances in our understanding
of low-mass X-ray binaries (LMXBs) thanks to precision-timing
observations made with the {\it Rossi X-ray Timing Explorer\/}
satellite (\rxte; \markcite{Bradt, Rothschild \& Swank 1993}).
Quasi-periodic oscillations (QPOs) in the 200--1200 Hz frequency range
have been detected in the persistent emission of many LMXBs.  These
QPOs often occur simultaneously separated by a frequency difference
that remains roughly constant or decreases as the frequencies
increase; these frequencies may reflect the relativistic motion of
material at the inner edge of the accretion disk in these systems.
For a recent review see \markcite{van der Klis (2000)}.

Separately, nearly-coherent oscillations (NCOs) with $Q$
($\nu/\Delta\nu$) values of $\gtrsim$300 have been detected during
many type~I X-ray bursts from six LMXBs, and there is strong evidence
that these frequencies reflect the underlying neutron star spin period
for these sources (\markcite{Strohmayer \etal\ 1996}; for a review see
\markcite{Strohmayer 1999}).  In particular, (1) the high coherence of
the signals is a challenge to alternative models: $Q > 900$ in a
single burst from KS~1731$-$260 (\markcite{Smith, Morgan, \& Bradt
1997}), and $Q$ values of \twid{4000} have been demonstrated, by
modeling of the frequency evolution during the burst, for bursts from
\slowb\ and 4U~1702$-$429 (\markcite{Strohmayer \& Markwardt 1999})
and \ksxrb\ (\markcite{Muno \etal\ 2000}); (2) the long-term stability
(to better than one part in 1000) of the NCO period in at least three
sources has been demonstrated over a time scale of $\sim$1 year
(\markcite{Strohmayer \etal\ 1998b}; \markcite{Muno \etal\ 2000}); (3)
the strength of the oscillations varies throughout the burst as
expected for a spin modulation: the oscillations are generally
strongest on the leading edge and tail of the bursts and undetectable
at the burst peak, when a large portion of the NS surface is in
conflagration (\markcite{Strohmayer, Swank \& Zhang 1998a}); and (4)
pulse phase-resolved spectroscopy of a burst from 4U~1636$-$536
indicates that the flux modulation during the tail of the burst is
accompanied by a modulation of the blackbody temperature of the
spectrum, again as expected (\markcite{Strohmayer \etal\ 1998a}).

The Rapid Burster (\mxbrb, or RB hereafter; \markcite{Lewin \etal\
1976}) is a recurrent transient LMXB, located in the Galactic plane at
a distance of approximately 8.6~kpc (\markcite{Frogel \etal\ 1995}) in
the highly reddened globular cluster Liller~1 (\markcite{Liller
1977}).  It is the only LMXB known to produce both type~I and type~II
X-ray bursts (\markcite{Hoffman, Marshall \& Lewin 1978}).  Whereas
the type~I bursts result from the explosive thermonuclear burning of
accreted material on the surface of a neutron star, the type~II bursts
result from spasmodic accretion -- the release of gravitational
potential energy -- presumably because of a recurrent accretion
instability. For a review, see \markcite{Lewin, van Paradijs, \& Taam
(1993)}.

\markcite{Guerriero (1998)} undertook a search for pulsations from the
type~I bursts, type~II bursts, and persistent emission of the Rapid
Burster, as observed through the first four outbursts of the source
seen with \rxte.  No significant detections were made, with upper
limits for persistent signals of $\approx$3 per cent RMS variation in
the range 20--2048 Hz.  Given the unique nature of the Rapid Burster,
we undertook to perform a more sensitive search.  We used the
knowledge that has been gleaned from other sources about the NCO
phenomenon, focusing our search on the leading edge of the type~I
bursts, and on the higher energy photons only (see below,
Sect.~\ref{sec:obs}).  Preliminary results of this work were reported
by \markcite{Fox \& Lewin (1999a)}.

\section{Observations and Analysis}
\label{sec:obs}
As of January~2000, seven outbursts of the Rapid Burster have been
observed with \rxtez\ (\markcite{Lewin \etal\ 1996a,b};
\markcite{Guerriero, Lewin \& Kommers 1997}; \markcite{Guerriero
\etal\ 1998}; \markcite{Fox \& Lewin 1998}; \markcite{Fox \etal\
1998}; \markcite{Fox \& Lewin 1999b}; the sixth outburst began in
March~1999 and was not reported).  \rxtez\ All-Sky Monitor (ASM;
\markcite{Levine \etal\ 1996}) data for these outbursts are available
in ``quicklook'' and fully-processed form from the web sites of the
ASM team\footnote{http://xte.mit.edu/XTE/asmlc/ASM.html} and the
High-Energy Astrophysics Science Archive Research
Center,\footnote{http://heasarc.gsfc.nasa.gov/docs/xte/asm\_products.html}
respectively.

Alerted by the ASM, we are able to commence target of opportunity
observations with the Proportional Counter Array (PCA;
\markcite{Jahoda \etal\ 1996}) within one day of the start of an
outburst, and to follow it throughout its month-long evolution
(\markcite{Guerriero \etal\ 1999}).

Outburst profiles, as observed with the \rxtez\ ASM, typically exhibit
a sharp 1--3~d rise to a maximum flux of $\approx$400~mCrab
(2--12~keV; the Crab produces 75~\ctsec\ in the ASM cameras), as
time-averaged over the 90-second ASM dwells.  The X-ray flux then
declines quasi-exponentially over the next month with a time constant
of $\approx$8~d (\markcite{Guerriero \etal\ 1999}).  Note, however,
that the sixth outburst in March~1999 deviated significantly from this
pattern (\markcite{Fox \etal\ 1999}).

The timing analysis presented here relies exclusively on the PCA data.
Our observations used individually described, event-encoded data with
a time resolution of 122~$\mu$s and 64 energy channels.  For the usual
spectral parameters of the RB, one PCA \ctsec\ $\approx 3\times
10^{-12}$ \ergcms, 2--20~keV (\markcite{Guerriero \etal\ 1999}).  When
the outburst reaches a maximum (PE and type~I bursts), count rates are
$\approx$5000 PCA \ctsec\, corresponding to an isotropic luminosity of
$1.3\times 10^{38}$ \ergsec\ (2--20 keV) at a distance of 8.6~kpc.
The brightest type~I burst peak luminosities, after PE subtraction,
are similarly \twid{$10^{38}$} \ergsec\ (see Table~1).
\begin{table*}
 \centering
 \begin{minipage}{100mm}
 \caption{Type I Bursts Observed by \rxte}
 \begin{tabular}{l@{~}llrrrrr}
 ~ & ~ & ~ & ~ & \multc{PE} & \multc{Peak} & \multc{Rise} & \multc{Merit} \\
 \multcc{2}{Burst} & 
 \multc{Date} &
 \multc{UT} & 
 \multc{(\ctsec)} &
 \multc{(\ctsec)} &
 \multc{(s)} &
 \multc{(c s$^{-2}$)} \\ \hline
\multcc{8}{Outburst 2} \\
\bf   1       &(a)& 1996-11-06 & 20:00:49 &  3256 & 4960 & 1.4 &\bf 2138 \\
\bf   2\Tfn{a}&(a)& 1996-11-09 & 01:06:20 &  1056 & 2664 & 1.3 &\bf 1496 \\
      3\Tfn{a}&(c?)&1996-11-11 & 21:52:45 &   816 &  656 & 0.2 & 1856 \\
\multcc{8}{Outburst 3} \\
      4      &(b)& 1997-06-26 & 05:03:43 &  5360 & 2072 & 2.4 &  240 \\
      5      &(b)& 1997-06-26 & 05:13:23 &  5288 & 2272 & 5.6 &  123 \\
      6\Tfn{a}&(b)& 1997-06-26 & 05:22:42 &  2112 & 1000 & 4.6 &   70 \\
      7      &(b)& 1997-06-26 & 08:21:13 &  5696 & 2368 & 2.8 &  251 \\
      8      &(b)& 1997-06-26 & 08:28:59 &  5784 & 1632 & 7.2 &   50 \\
      9\Tfn{a}&(b)& 1997-06-26 & 08:40:36 &  2368 & 1144 & 2.6 &  145 \\
     10\Tfn{a}&(b)& 1997-06-26 & 08:48:51 &  2200 & 1120 & 3.6 &  105 \\
\bf  11      &(b)& 1997-06-27 & 18:13:15 &  5152 & 3448 & 2.2 &  642 \\
\bf  12\Tfn{b}&(b)& 1997-06-27 & 18:30:34 &  5264 & 2768 & 1.8 &  536 \\
\bf  13\Tfn{a}&(a)& 1997-06-29 & 07:03:31 &  1496 & 2016 & 1.5 &  783 \\
\bf  14      &(a)& 1997-06-29 & 18:07:23 &  4016 & 4000 & 1.4 & 1409 \\
\bf  15\Tfn{a}&(a)& 1997-06-29 & 18:35:24 &  1656 & 1560 & 0.6 & 1226 \\
\bf  16\Tfn{a}&(a)& 1997-07-07 & 13:29:06 &   664 & 2088 & 1.1 & 1391 \\
\bf  17\Tfn{a,b}&(a)& 1997-07-10 & 14:03:41 &   528 & 2208 & 1.2 &\bf 1473 \\
\bf  18      &(a)& 1997-07-13 & 12:01:19 &  1136 & 1728 & 1.5 & 706 \\
\multcc{8}{Outburst 4} \\
\bf  19      &(a)& 1998-01-30 & 18:31:04 &  4680 & 3424 & 2.0 &  716 \\
\bf  20\Tfn{b}&(a)& 1998-01-30 & 18:54:18 &  4784 & 3424 & 1.6 &  867 \\
\bf  21      &(a)& 1998-01-30 & 20:02:22 &  4344 & 3584 & 1.5 & 1088 \\
\bf  22      &(a)& 1998-01-30 & 20:28:12 &  4328 & 3864 & 1.7 & 1082 \\
     23\Tfn{c}&(a)& 1998-01-30 & 20:54:18 &  4232 & 4200 & 1.5 & n/a \\
\bf  24      &(a)& 1998-01-31 & 23:25:27 &  3672 & 3664 & 1.6 & 1169 \\
\bf  25      &(a)& 1998-01-31 & 23:53:09 &  3712 & 3736 & 1.3 &\bf 1494 \\
\bf  26      &(a)& 1998-02-02 & 17:00:47 &  2480 & 4720 & 1.3 &\bf 2349 \\
\bf  27      &(a)& 1998-02-02 & 18:46:18 &  2552 & 4912 & 1.2 &\bf 2703 \\
\bf  28      &(a)& 1998-02-04 & 20:21:48 &  2288 & 4960 & 1.7 &\bf 1954 \\
\bf  29      &(a)& 1998-02-07 & 19:00:34 &  1776 & 5328 & 1.5 &\bf 2746 \\
\bf  30      &(a)& 1998-02-07 & 20:04:54 &  1744 & 4920 & 1.3 &\bf 2800 \\
\bf  31\Tfn{g}&(c)& 1998-02-10 & 22:06:44 &  1424 & 4208 & 1.2 &\bf 2587 \\
\bf  32      &(c)& 1998-02-16 & 14:04:17 &  1256 & 3632 & 0.5 &\bf 5292 \\
     33\Tfn{d}&(c)& 1998-02-16 & 15:13:01 &  1168 & 1608 & 2.3 & n/a \\
\bf  34\Tfn{a}&(c)& 1998-02-19 & 14:29:55 &   192 & 2896 & 0.3 &\bf 8103 \\
\multcc{8}{Outburst 5} \\
\bf  35\Tfn{e,g}&(a?)& 1998-08-19 & 05:07:19 & 13664 & 5232 & 4.7 &  306 \\
\bf  36\Tfn{e} &(a?)& 1998-08-20 & 13:21:38 & 10848 & 3256 & 1.0 &  765 \\
\bf  37\Tfn{f} &(d)& 1998-08-22 & 08:24:11 &  3328 & 3448 & 1.3 & 1313 \\
\bf  38\Tfn{f} &(d)& 1998-08-22 & 09:01:59 &  3272 & 3384 & 1.6 & 1107 \\
\bf  39\Tfn{a,f}&(d)& 1998-08-22 & 10:15:45 &  1184 & 1384 & 1.6 &  453 \\
\bf  40        &(c)& 1998-09-01 & 02:13:16 &  1152 & 4480 & 0.6 &\bf 5603 \\
\bf 42x\Tfn{f} &(c?)& 1998-09-07 & 03:35:12 &   608 & 2288 & 3.3 & 541 \\
\multcc{8}{Outburst 6} \\
     41\Tfn{g}&(a)& 1999-03-12 & 21:50:02 &  3440 & 2776 & 2.7 &  467 \\
     42\Tfn{g}&(a)& 1999-03-12 & 22:12:53 &  3344 & 2928 & 1.5 &  935 \\
     43\Tfn{i}&(a)& 1999-03-16 & 02:32:05 &  1616 & 1488 & 1.2 &  593 \\
     44\Tfn{i}&(a)& 1999-03-16 & 03:00:04 &  1576 & 1448 & 1.4 &  501 \\
     45\Tfn{g}&(a)& 1999-03-16 & 03:55:32 &  3152 & 2600 & 0.9 & 1336 \\
     46\Tfn{g}&(a)& 1999-03-16 & 04:23:23 &  3104 & 2968 & 1.0 & 1411 \\
     47\Tfn{g}&(a)& 1999-03-19 & 19:57:22 &  1056 & 1448 & 1.4 &  586
 \\ \hline
 \end{tabular}
 \end{minipage}
\end{table*}

\begin{table*}
 \centering
 \begin{minipage}{100mm}
 \setcounter{table}{0}
 \caption{Type I Bursts Observed by \rxte\ (continued)}
 \begin{tabular}{l@{~}llrrrrr}
 ~ & ~ & ~ & ~ & \multc{PE} & \multc{Peak} & \multc{Rise} & \multc{Merit} \\
 \multcc{2}{Burst} & 
 \multc{Date} &
 \multc{UT} & 
 \multc{(\ctsec)} &
 \multc{(\ctsec)} &
 \multc{(s)} &
 \multc{(c s$^{-2}$)} \\ \hline
\multcc{8}{Outburst 7} \\
     48\Tfn{h} &(b)& 1999-09-30 & 20:35:00 &  2632 & 1848 & 1.6 &  466 \\
     49\Tfn{h} &(b)& 1999-09-30 & 20:55:54 &  2512 & 1880 & 2.7 &  299 \\
     50\Tfn{h} &(b)& 1999-09-30 & 21:17:23 &  2464 & 2048 & 2.2 &  420 \\
     51\Tfn{h} &(b)& 1999-09-30 & 22:03:53 &  2488 & 2136 & 1.6 &  614 \\
     52\Tfn{h} &(a)& 1999-09-30 & 22:26:54 &  2368 & 2408 & 2.0 &  607 \\
     53\Tfn{i} &(a)& 1999-10-02 & 09:28:25 &  1520 & 1728 & 1.8 &  518 \\
     54\Tfn{i} &(a)& 1999-10-02 & 09:54:57 &  1528 & 1896 & 1.8 &  572 \\
     55\Tfn{i} &(a)& 1999-10-02 & 10:49:25 &  1528 & 1744 & 1.6 &  565 \\
     56\Tfn{i} &(a)& 1999-10-02 & 11:16:28 &  1520 & 1760 & 1.6 &  596 \\
     57\Tfn{a,b,i}&(a)&1999-10-02 & 11:44:19 &   544 &  736 & 1.3 &  335 \\
     58\Tfn{d,h}&(a)& 1999-10-05 & 12:11:53 &  1752 & 3112 & 1.6 & 1273 \\
     59\Tfn{h}&(a)& 1999-10-05 & 12:49:39 &  1768 & 3160 & 1.7 & 1193 \\
     60       &(a)& 1999-10-08 & 12:40:07 &  2736 & 4672 & 1.3 &\bf 2335 \\
     61\Tfn{g}&(a)& 1999-10-12 & 12:29:52 &  1664 & 4024 & 1.0 &\bf 2926 \\
     62\Tfn{a,g}&(a)& 1999-10-12 & 14:28:52 &   632 & 1744 & 1.6 &  825 \\
     63        &(a)& 1999-10-16 & 04:13:44 &  1344 & 4688 & 1.0 &\bf 3487 \\
     64\Tfn{a,i}&(c)& 1999-10-19 & 06:12:23 &   176 & 1872 & 0.2 &\bf 8612 
\\ \hline
 \end{tabular}
 \medskip

 Dates and times are in UT.  Profile classes for the bursts correspond
 to the examples shown in Fig.~\ref{fig:profiles}.  Persistent
 emission + background (PE) and burst peak count rates (Peak; for
 1/8-second bins) are in PCA cts/sec (2--60 keV). Burst rise times (in
 seconds) are characteristic times for exponential fits; Merit figures
 are calculated as Merit = (Peak$^2$/PE)/Rise.  Boldface burst numbers
 indicate the 31~bursts of the original search; boldface Merit figures
 indicate the 17~bursts of the final search; see text for details.
 \medskip

 \Tfn{a}{Observed during offset pointing (see text)}; \Tfn{b}{Burst
 tail truncated by slew or data gap}; \Tfn{c}{Not recorded in Event
 Mode data}; \Tfn{d}{Only partially recorded in Event Mode data};
 \Tfn{e}{Occurred during a type II burst}; \Tfn{f}{Identification as
 type I burst is uncertain}; \Tfn{g}{4 of 5 PCUs operational};
 \Tfn{h}{3 of 5 PCUs operational}; \Tfn{i}{2 of 5 PCUs operational}

\end{minipage}
\end{table*}

Type~I X-ray bursts are distinguished from the type~II bursts of the
RB in that they show a pronounced cooling over the course of the
burst; type~II bursts, on the other hand, maintain a roughly constant
temperature through the burst.  Near the maximum of the RB outbursts,
when the persistent emission is at its strongest, we have occasionally
observed bursts with a slow rise, slow decay and relatively low peak
flux -- classified as b-profile bursts below -- that we have
difficulty identifying definitively as type~I or type~II.  For the
most part we have identified them as type~I on the basis of spectral
analyses (\markcite{Guerriero \etal\ 1999}); however, this
identification is not certain in all cases.

\begin{figure*}
\centerline{~\psfig{file=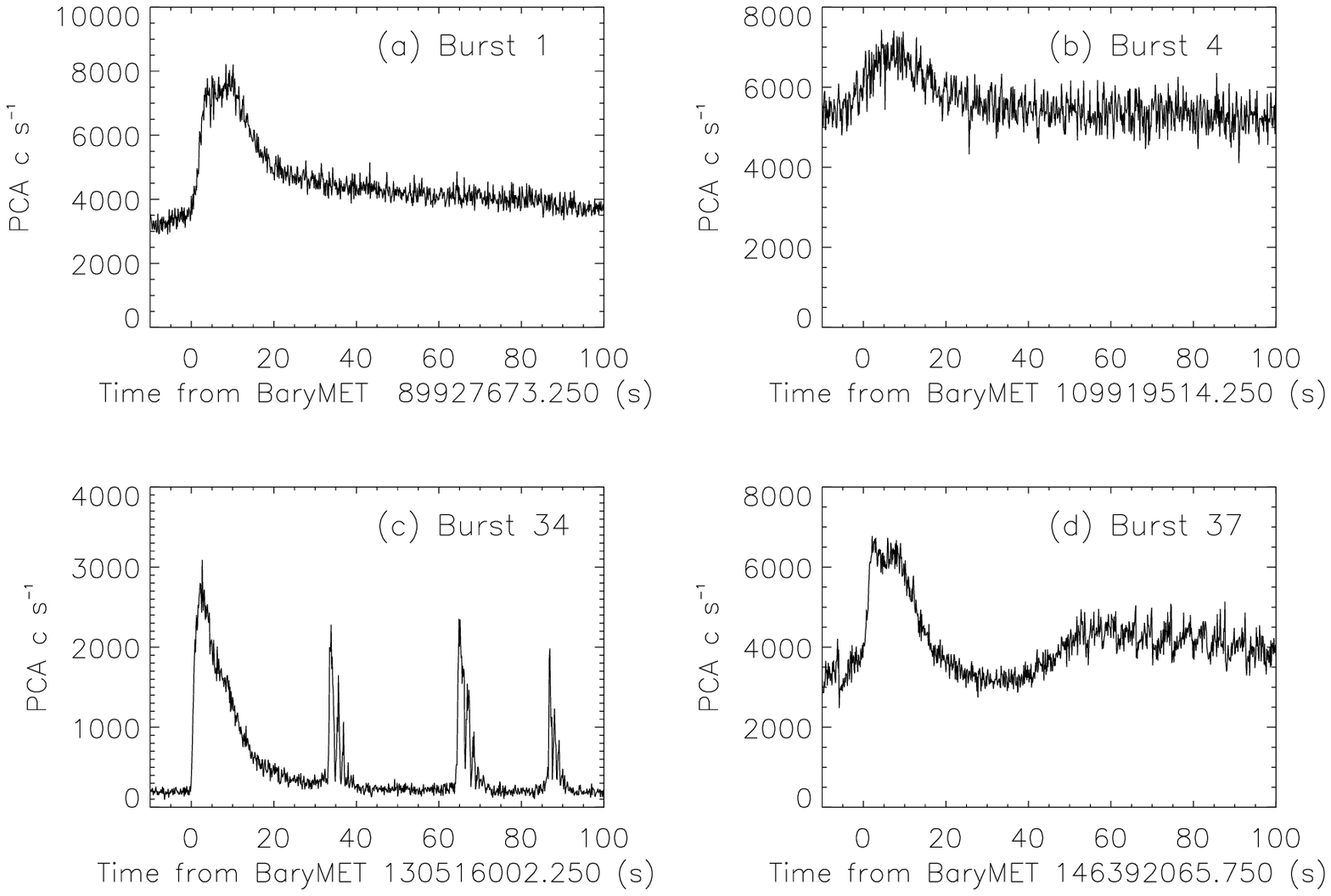,width=5.0in}~}
\caption{Characteristic profiles of different varieties of type~I
burst from the Rapid Burster, as seen by the \rxtez\ PCA (raw count
rates only): (a) Two-phase decay, with bursts lasting $\gtrsim$100~s;
(b) Slow rise and symmetric decay; (c) Fast rise and one-phase decay
-- note that three type~II bursts are also visible in this panel; (d)
type~I burst followed by increased PE with very strong QPO -- the
three bursts in this category (\#37-39) were observed on 22~August
1998 and will be the subject of a separate paper.  These are the four
burst profile classes indicated in Table~1. }
\label{fig:profiles}
\end{figure*}

In Figure~\ref{fig:profiles} we present the profiles of four type~I
bursts from the Rapid Burster that indicate the variety of burst
profiles that we observe.  Table~1 identifies all of the bursts by
profile class; note, however, that these identifications are merely
gross characterizations, and in particular, the distinction between
profile classes (a) and (b) is probably not well-defined at this
point.

We note that an occasional type~I burst from \slowb\ is visible at
$\approx$50 per cent collimation efficiency during direct pointings at
the RB, but that these bursts exhibit spectra (fitted blackbody
temperatures ($\approx$2~keV), peak count rates ($\approx$10,000
\ctsec), and light curves ($\approx$10~sec decays) that readily
distinguish them from the RB bursts.

Our analysis evolved historically and must be described as such. It
began in 1998 after \rxtez\ had observed five outbursts from the Rapid
Burster.  We compiled a list of all RB type~I bursts that were
observed with \rxtez\ during these outbursts (Table~1, data through
September~1998); note that no type~I bursts were observed during the
first outburst (April and May 1996). We selected one-second time
segments from the start of each burst: an automated procedure found
the first eighth-second bin to represent a 4.5$\sigma$ fluctuation
above the background level, and these start times were then inspected
visually and adjusted by a half-second or less by hand.

To maximize our sensitivity to pulsations modulating the burst flux
alone (and not the background or PE), we rank-ordered the bursts in
terms of Peak$^2$/PE (Table~1), and selected the top 31 bursts for our
sample, as these seemed the most attrative targets; the selected
bursts are indicated in bold in Table~1.  To increase signal-to-noise
for a relatively hard pulsation we selected photons in the 5.5--16~keV
bandpass, roughly the upper half of the energy range for the Rapid
Burster spectrum.

We binned the light curves from these one-second intervals to
$2^{-12}$~sec time resolution, corrected times to the solar system
barycentre, took the Fourier transform, and averaged the 31 resulting
power density spectra -- producing a single power density spectrum
(PDS) of 1~Hz frequency resolution and Nyquist frequency 2048~Hz.

\subsection{Search Procedure}
\label{sub:srch}
The frequency of the NCOs often drifts by roughly 2~Hz during the
bursts (\markcite{Strohmayer \etal\ 1996}), so we do not necessarily
anticipate a one-bin excess in the PDS.  Rather than search for the
highest individual powers, then, we began by inspecting the running
two-bin average of adjacent powers from the PDS.  Then, in a check for
broader signals, we rebinned the PDS by successive factors of two and,
at each stage, performed a two-bin running-average search on the
averaged powers.

Since the statistical distribution of Poisson noise powers is known
(for raw powers it is the $\chi^2$ distribution with two degrees of
freedom; \markcite{Leahy \etal\ 1983}), we were able to compare the
search results from different stages of this process and select those
which are most significant in a global sense.  In fact, averaging 31
PDSs already puts the powers in a near-Gaussian regime, and we adopted
this approximation in ranking prospective triggers of different bin
sizes.

We have performed Monte Carlo calculations to determine the absolute
significance of any particular excess over the entire power-spectral
search.  We did this because, first, the running-average search
procedure (with its multiple rebinnings and reexaminations of the PDS)
results in a distribution of tested powers that does not have an
analytical form; and second, because the burst rise injects strong
low-frequency power and places sidebands on noise spikes, increasing
the noise powers for an integral-power search such as ours.

\subsection{Monte Carlo Method}
\label{sub:mc}
Our Monte Carlo (MC) procedure used the actual burst data, scrambling
the data on short time scales to destroy real high-frequency power
while preserving the burst rise and low-frequency noise responsible
for the amplification of noise spikes in the PDS.  The 120 time bins
from each 30~msec interval of a burst were randomly interchanged prior
to performing the FFT.  We then averaged the 31 ``fake'' PDSs from the
individual bursts and performed our full search protocol, recording
the highest three signals that were found.  For the purposes of this
paper, we performed an MC run of 100,000 trials.

There are two aspects of this procedure which, if ignored, will
systematically affect the resulting statistics.  First, real noise is
present in our fake PDSs at low frequencies, $\lesssim$100~Hz, because
of the nature of our scrambling procedure.  We therefore ignore false
triggers in the frequency range $<$60~Hz.

Second, the MC procedure destroys real broad-band power at high
frequencies, which (if not accounted for) will systematically enhance
powers in the real PDS relative to powers in the fake (MC) ones.  We
have therefore characterized the broad-band excess power that is
present in the real data. Subtracting off the deadtime-corrected PCA
Poisson level (\markcite{Morgan, Remillard \& Greiner 1997}) from our
real (averaged) PDS, and logarithmically rebinning, we find that we
may fit the residuals with a power law of exponent $-$0.7 and total
normalization 1.0 per cent rms over 1--2048~Hz.  We corrected our real
PDS powers by dividing by this background level before comparing our
powers to the MC results.

\subsection{Candidate Signal Identification}
\label{sub:sig}
\begin{figure}
\centerline{~\psfig{file=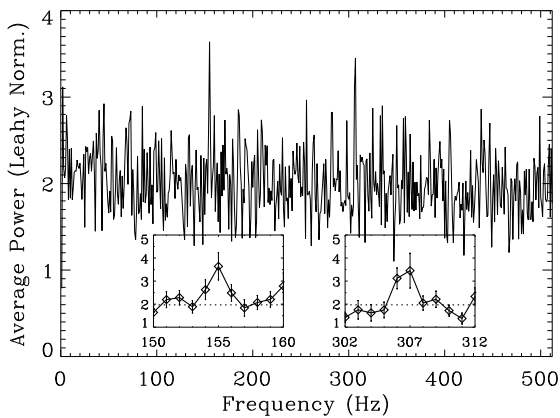,width=3.4in}~}
\caption{Average power density spectrum of one second of data from the
leading edge of 31 type~I X-ray bursts (boldface in Table~1), energy
range 5.5--16 keV.  Powers in the vicinity of the frequencies 155~Hz
and 306.5~Hz are shown in close-up views.  Error bars in the close-up
views are the sample standard deviations of the averaged points.  Note
that although the single-bin 155~Hz power is greater than the 306~Hz
and 307~Hz powers, our search procedure finds the 306.5~Hz candidate
signal to be of greater significance (see text for details).}
\label{fig:discovery}
\end{figure}

The average power density spectrum of the first one second of the 31
type~I bursts (see Table 1) from the RB is presented in Fig.~\ref{fig:discovery}.
The two highest peaks, centred at 306.5~Hz and 155~Hz, respectively,
extend noticeably above the noise level, and are placed in a
near-harmonic relationship.  Specifically, our Monte Carlo results
indicate that a peak with the strength of the 306.5~Hz peak (or
stronger) is present in 1.8 per cent of the simulated power density
spectra, and that a peak with the strength of the 155~Hz peak, or
stronger, is present in 15.3 per cent of the simulated PDSs.  Note
that although the single-bin 155~Hz power is the strongest in the PDS,
the 306.5~Hz candidate signal is more significant because of the
strong power in two adjacent bins.  The chances of any PDS having two
peaks, both stronger than the strength of our 155~Hz peak, is 1.2 per
cent.

The near-harmonic frequencies of the two peaks may be evidence for a
common physical origin.  However, the 1 per cent difference in
frequencies is uncomfortably large to result from the frequency drifts
that have been observed in type~I bursts from other sources
(\markcite{Strohmayer \& Markwardt 1999}; \markcite{Muno \etal\ 2000})
-- though a connection can certainly not be excluded.  The burst
frequencies in other sources drift by less than 1 per cent, total, and
exhibit the most pronounced evolution during the tails of the bursts.
Moreover, the signal we observe here, if real, results from the
contribution of several bursts, which would tend to diminish the
observable effects of the frequency drift.

In the event that the two signals are related then their combined
significance is still dependent upon {\it post factum} assumptions; we
give an illustrative example. If we allow for a $\pm$3 per cent shift
in the frequency of a harmonic or subharmonic relative to the main
signal (conservative in that the actual shift is 1.1 per cent) then we
would consider two frequency windows, 153--158~Hz and 594--632~Hz --
centred on one-half and double the main 306.5~Hz signal frequency,
respectively -- as candidate windows for harmonic detection.  The
chance of observing a signal as strong as our 155~Hz signal (or
stronger) within this range is roughly 0.38 per cent (15.3 per cent
full-search probability multiplied by the size of this restricted
range and divided by the size of the full 60--2048~Hz window).
Multiplying the odds for the main signal (54:1) by the odds for this
additional signal (259:1) would give a combined probability for the
dual detection of $7.0\times 10^{-5}$, corresponding to a 4.0$\sigma$
level of confidence.

\subsection{Outbursts 6 and 7}

After we had finished the analysis of the type~I bursts of the first
five outbursts as described above, and before we were ready to submit
our results for publication, two additional outbursts of the RB
occurred in March and September 1999, respectively.  The few bursts
from Outburst~6 were of generally low quality, but during Outburst~7
better bursts were collected and so we performed a full analysis on
those 23~bursts (see Table~1).  We found no evidence for modulation
near 306~Hz, or at any other frequency. We can set 95 per cent
confidence RMS upper limits of 8.1 per cent, 7.7 per cent, and 5.0 per
cent, respectively, on the average 306 + 307~Hz power during the one
second before the rise, the one second rise, and the one second after
that.

We then analyzed all 61 type~I bursts with fast-timing data (covering
the first seven {\it RXTE} outbursts).  As with our initial search, we
found the strongest signal in the 306 + 307~Hz bin.  The RMS variation
in the average PDS is 7.0$\pm$1.1 per cent during the first one second
of the bursts (5.5--16~keV energy band).  Monte Carlo simulations
indicate, however, that this signal is not significant: 19 per cent of
the simulated datasets have signals of equal or greater significance.
If the signal is not real, then it sets our quoted 95 per cent
confidence upper limit for a narrow pulsation: 8.8 per cent rms on
average.  95 per cent confidence upper limits on the average
306 + 307~Hz RMS modulation in the one second before and the one second
after the first second, for the full burst set, are 5.7 per cent and
2.2 per cent, respectively.

\subsection{Candidate Signal Follow-Up}

Cumming \& Bildsten (2000) have shown that the vertical thermal time
during mixed hydrogen/helium burning is longer than the wrapping time
from the expansion-induced differential rotation, and that the
opposite is true for helium-rich bursts: the heat from these bursts
can get out faster and, in addition, the shearing frequency is
lower. This makes it easier for a periodic signal from some underlying
asymmetry to escape from a helium-rich burst. Muno \etal\ (2000) have
found that the ``fast'' bursts from \ksxrb\ -- bursts with high peak
fluxes, short decay times, and radius expansion, indicative of helium
burning -- occur in the same part of the X-ray colour-colour diagram,
and all exhibit burst pulsations. The ``slow'' bursts, likely the
result of mixed hydrogen/helium burning -- with lower peak fluxes,
longer decay times, and no radius expansion -- do not show pulsations,
consistent with this picture. Lewin \etal\ (1987) showed that the
bursts from 1636$-$53 with radius expansion had substantially shorter
rise times than those that did not exhibit radius expansion.

Inspired by the above, we restricted our attention to a set of 17
fast-rise bursts from all seven outbursts, and we recomputed an
average PDS, and performed our signal search again.  These bursts were
selected on the basis of their ``Merit'' function (see Table~1),
defined as (Peak$^2$/PE)/Rise to take into account the expected
strength of any burst signal, as well as the distinction between fast-
and slow-rise bursts.  Note however that we have no particular
evidence that these bursts (or any bursts from the RB) resulted from
helium-rich as opposed to hydrogen-rich burning; this question will be
addressed in future work.  We ignore Burst~3, a burst with low
signal-to-noise that has high ``Merit'' only because of its unusually
short rise time.

We find that the signal is strengthened significantly, with an average
power in this burst sample of 4.13 for the two-bin 306.5~Hz average
(mean of 34 powers).  A signal of this strength or stronger is found
in only 0.07 per cent of the Monte Carlo simulations of this dataset,
corresponding to a 3.4-sigma level of confidence.  The strength of the
signal corresponds to an average 8.0 per cent RMS modulation of the burst
flux during these intervals.

Similar searches of the burst flux preceding and following the
selected one second do not show significant signal power; the 95 per
cent confidence upper limit on the excess average 306 + 307 Hz power
during these intervals is 0.76 and 0.47, respectively.  These
correspond to average upper limits of 10 per cent and 2 per cent RMS,
respectively, for modulation of the burst flux during these intervals.

Likewise, a search for power in the soft band (2.0--5.5~keV) photons
from the first second of each burst does not show significant power:
the 95 per cent confidence upper limit on the excess average 306 + 307 Hz
power during these intervals is 0.69, corresponding to an average RMS
upper limit of 4.3 per cent for modulation of the soft burst flux.

We performed a separate systematic search for pulsations near 307~Hz
during the tails of all type~I bursts (except for Burst~3, which is
too weak for useful limits to be set).  The search parameters here
were chosen to maximize signal for the relatively weaker pulsations
during the burst tails: we select higher-energy photons only
(5.5--16~keV), divide the first 16~s of each burst into four 4~s
segments, FFT and search for the highest power in the 304--310~Hz
range.  We found no significant power, and our mean per-burst 95 per
cent confidence upper limits on the strength of any coherent power
(0.25~Hz bins) in the tails of the bursts are 5 per cent--14 per cent
RMS, relative to the 5.5--16~keV burst flux, depending on the strength
of the burst.  Given the weaker signal strengths that are generally
observed during burst tails, and the large search space we necessarily
considered, we do not consider this non-detection surprising.

Since our only finding, subsequent to the evidence uncovered in our
initial search, was a strengthening of that evidence within an
overlapping set of bursts, this result is subject to a posteriori
effects and is of uncertain absolute significance.

\section{Conclusions}

We have performed a search for burst pulsations from the type~I bursts
of the Rapid Burster.  The search was targeted to maximize the chances
for detection of a signal like those found in other type~I burst
sources, where the maximum modulation amplitude occurs at the leading
edge of a burst and the pulsations are spectrally hard compared to the
burst and persistent emission.  We find evidence at the $>$98 per cent
confidence level for a modulation near 307~Hz which, if confirmed,
will indicate a spin period for the Rapid Burster of 3.25 or 6.5
milliseconds, depending on whether the main burst signal is the
fundamental or first harmonic of the spin frequency.

\section*{Acknowledgments}

We thank Deepto Chakrabarty, Tod Strohmayer and Fred Lamb for helpful
discussions.  W.L. gratefully acknowledges NASA support.  L.B. was
partially supported by NASA via grant NAG 5-8658 and by the National
Science Foundation under grant no.\ PHY 94-07194.  L.B. is a Cottrell
Scholar of the Research Corporation.

\label{lastpage}

\end{document}